%% file: main.tex
\documentclass[sigconf]{acmart}

\usepackage{booktabs}
\usepackage{multirow}
\usepackage{enumitem}
\usepackage[most]{tcolorbox}
\usepackage{tabularx}

\AtBeginDocument{%
  \providecommand\BibTeX{{%
    Bib\TeX}}}

\copyrightyear{2026}
\acmYear{2026}
\setcopyright{cc}
\setcctype{by}
\acmISBN{979-x-xxxx-xxxx-x/YYYY/MM}
\acmDOI{10.1145/xxxxxxx.xxxxxxx}
\begin{document}

\title{Unbiased Multimodal Reranking for Long-Tail Short-Video Search}

% \author{Wenyi Xu}
% \affiliation{%
%   \institution{Zhejiang University; Kuaishou Technology}
%   \city{Hangzhou}
%   \country{China}
% }
% \email{xuwenyi@zju.edu.cn}

\author{Wenyi Xu}
\affiliation{%
  \institution{Zhejiang University}
  \state{Hangzhou}
  \country{China}
  }
\affiliation{
  \institution{Kuaishou Technology}
  \city{Hangzhou}
  \country{China}
}
\email{xuwenyi@zju.edu.cn}

\author{Feiran Zhu}
\authornote{Corresponding author.} 
\affiliation{%
  \institution{Kuaishou Technology}
  \state{Hangzhou}
  \country{China}}
\email{zhufeiran03@kuaishou.com}

\author{Songyang Li}
\affiliation{%
  \institution{Kuaishou Technology}
  \state{Beijing}
  \country{China}}
\email{lisongyang03@kuaishou.com}

\author{Renzhe Zhou}
\affiliation{%
  \institution{Kuaishou Technology}
  \state{Hangzhou}
  \country{China}}
\email{zhourenzhe03@kuaishou.com}

\author{Chao Zhang}
\affiliation{%
  \institution{Kuaishou Technology}
  \state{Hangzhou}
  \country{China}}
\email{zhangchao29@kuaishou.com}

\author{Chenglei Dai}
% \authornote{Corresponding author.} 
\affiliation{%
  \institution{Kuaishou Technology}
  \state{Hangzhou}
  \country{China}}
\email{daichenglei@kuaishou.com}

\author{Yuren Mao}
\affiliation{%
  \institution{Zhejiang University}
  \state{Hangzhou}
  \country{China}}
\email{yuren.mao@zju.edu.cn}

\author{Yunjun Gao}
\affiliation{%
  \institution{Zhejiang University}
  \state{Hangzhou}
  \country{China}}
\email{gaoyj@zju.edu.cn}

\author{Yi Zhang}
\affiliation{%
  \institution{Kuaishou Technology}
  \state{Beijing}
  \country{China}}
\email{zhangyi49@kuaishou.com}

\renewcommand{\shortauthors}{Wenyi Xu et al.}

\begin{abstract}
Kuaishou serving hundreds of millions of searches daily, the quality of short-video search is paramount. However, it suffers from a severe Matthew effect on long-tail queries: sparse user behavior data causes models to amplify low-quality content such as clickbait and shallow content. 
The recent advancements in Large Language Models (LLMs) offer a new paradigm, as their inherent world knowledge provides a powerful mechanism to assess content quality, agnostic to sparse user interactions.
To this end, we propose a LLM-driven multimodal reranking framework, which estimates user experience without real user behavior. The approach involves a two-stage training process: the first stage uses multimodal evidence to construct high-quality annotations for supervised fine-tuning, while the second stage incorporates pairwise preference optimization to help the model learn partial orderings among candidates. At inference time, the resulting experience scores are used to promote high-quality but underexposed videos in reranking, and further guide page-level optimization through reinforcement learning. Experiments show that the proposed method achieves consistent improvements over strong baselines in offline metrics including AUC, NDCG@K, and human preference judgement. An online A/B test covering 15\% of traffic further demonstrates gains in both user experience and consumption metrics, confirming the practical value of the approach in long-tail video search scenarios.
\end{abstract}

%%
%% The code below is generated by the tool at http://dl.acm.org/ccs.cfm.
%% Please copy and paste the code instead of the example below.
%%




% \received{20 February 2007}
% \received[revised]{12 March 2009}
% \received[accepted]{5 June 2009}

\maketitle

\section{Introduction}
\input{main_text/introduction}

\section{Related works}

\input{main_text/related_work}

\section{Methodology}

\input{main_text/method}
\section{Experiments}
\input{main_text/experiments}

\section{Conclusion}
\input{main_text/conclusion}

\bibliographystyle{ACM-Reference-Format}
\bibliography{sample-base}

\appendix
\input{main_text/appen}

\end{document}

%% file: main_text/introduction.tex
On leading short-video platforms serving hundreds of millions of searches daily, search quality is paramount. While long-tail queries constitute a significant fraction of this volume—approximately 20\%—they are disproportionately responsible for poor user experiences. For these queries, the scarcity of user interaction data creates a vicious cycle known as the Matthew effect~\citep{Popularity_bias}. Lacking reliable signals, ranking models often over-rely on superficial cues like sensational thumbnails or catchy titles. Consequently, the results page is frequently dominated by clickbait and shallow content, which steadily erodes user trust and platform reputation~\citep{click_cheating}.

In the long-tail regime, sparse behavioral signals make ranking especially vulnerable to three entangled biases. Position/exposure bias amplifies items merely because they were seen rather than truly preferred, leading models to overfit presentation effects. Popularity bias further distorts supervision by rewarding click-bait and over-exposed creators, especially when feedback is scarce. Finally, cross-modal mismatch, where the cover/title narrative diverges from the actual footage, obscures intrinsic quality. These effects are exacerbated by video’s multimodal nature: relevance must be inferred from titles, ASR transcripts, OCR snippets, covers, and key frames, making it hard to disentangle genuine usefulness from noise under weak feedback ~\citep{ThumbnailTruth, Localizing}.

Existing remedies are insufficient as they fail to address these coupled challenges: multimodal inconsistency and query-specific relevance. On one hand, methods like query rewriting operate only at the textual level. They are fundamentally blind to the multimodal nature of video and thus cannot detect critical inconsistencies between a video's title, its spoken content (ASR), and its actual visual footage~\citep{taobao_rewritte}. On the other hand, while a generic video quality score might assess a video's internal coherence, its assessment is fundamentally query-agnostic. A well-produced video with perfect title-content consistency can still be completely irrelevant or even misleading for a specific long-tail query intent, making this signal unreliable for fine-grained ranking\citep{rec_overview, ugc}.

To address the three sources of long-tail degradation, we propose a unified solution. For popularity bias and cross-modal mismatch, we employ a Large Language Model with world knowledge to perform explicitly query-aware multimodal parsing, integrating titles, ASR, OCR, covers, and key frames. The model assesses intent–content consistency and intrinsic content value, based on which we conduct labeling and training to produce query-specific, comparable scores, thereby suppressing spurious strong signals such as clickbait, shallow content, and over-exposure at the source. For position/exposure bias, we anchor on a user-behavior-agnostic experience score to reconstruct reranking training signals and the page-level reinforcement-learning reward: at the point level, we reconstruct behavior sequences with the experience score to attenuate position effects; 
at the page level, we construct nDCG-style returns from the experience-induced ideal order, refining the reward function used in reinforcement learning, 
thus correcting exposure-induced systematic bias. 
With this design, without relying on user behavior, we ground multimodal consistency and query semantics into a deployable, lightweight reranking signal that mitigates long-tail biases and improves user experience.

The main contributions of this work are listed as follows:
\begin{itemize}[leftmargin=2.5em, labelsep=0.5em, itemsep=2pt, topsep=2pt]
  \item We analyze the three sources of bias in long-tail queries and propose a user-behavior-agnostic, explicitly query-aware multimodal experience-scoring model.
  \item We use the model’s experience scores to drive label reconstruction and page-level reward optimization in reranking.
  \item Offline studies and large-scale online A/B tests show stable improvements in user-experience metrics without sacrificing key consumption metrics.
\end{itemize}

%% file: main_text/related_work.tex
\subsection{Architecture for Reranking}

Modern search systems typically follow a two-stage cascade: an initial retrieval step rapidly recalls candidate documents, and a subsequent reranking step refines these candidates with more sophisticated models \cite{nogueira2019multistage,jacob2025drowning}.
Within this reranking stage, recent studies have explored large language model approaches from four viewpoints: pointwise, pairwise, listwise, and setwise\cite{abdallah2025good}. 
RankGPT\cite{sun2023chatgpt} casts reranking as a sequence-generation problem and already outperforms traditional baselines in zero-shot settings, while setwise prompting scores\cite{podolak2025beyond,zhuang2024setwise} an entire group of candidates in a single pass, markedly lowering inference cost without sacrificing effectiveness.
To remedy the score-incomparability issue of listwise outputs, Self-Calibrated Listwise Reranking\cite{ren2025self} introduces a “list view + point view” dual relevance scheme, using the point-view scores to calibrate listwise results and achieve global consistency. The open-source model RankVicuna\cite{pradeep2023rankvicuna} (7B parameters) further shows that even a relatively small LLM can reach GPT-3.5-level reranking quality in a zero-shot scenario. 
In addition, novel loss functions have been proposed: Softmax-DPO\cite{chen2024softmax} couples a Plackett–Luce softmax with multiple negatives, diffNDCG\cite{zhou2024optimizing} directly optimizes ranking metrics such as NDCG, and reinforcement-learning schemes like GRPO align LLM predictions with human preference signals\cite{zhuang2025rank}.
However, these methods generally assume that labeled data are both plentiful and unbiased; when the supervision itself is noisy or skewed, the advantages of LLM-based rerankers can diminish substantially.

\subsection{Personalized Ranking with LLMs}

In personalized ranking, researchers focus on injecting user interests and behaviors into LLM-based rankers.
For example, PREMIUM \cite{sun2025premium} encodes preferences via a tag system and lets users iteratively rank model outputs to achieve on-device, personalized fine-tuning of an LLM. 
For multimodal recommendation, NoteLLM-2\cite{zhang2025notellm} enriches item representations with image-and-text inputs so the LLM can capture visual preferences. 
CHIME\cite{bai2025chime} encodes a user’s entire behavior sequence with an adaptive LLM and, through contrastive learning plus quantization, produces compact long-term interest vectors for holistic preference modeling. 
HLLM\cite{chen2024hllm} adopts a hierarchical design: one LLM extracts content features from item descriptions, while another predicts future interests from historical interactions, thereby leveraging pre-trained knowledge in sequential recommendation. 
Work on online feedback loops is emerging as well: Wang\cite{wang2025user} decouple novelty and preference with two separate LLMs, then filter novelty-driven recommendations through a preference-aligned model to balance diversity and relevance, significantly boosting user satisfaction and diversity. 
Collectively, these studies demonstrate how LLMs can learn and adapt to behavioral data to deliver more personalized rankings. However, despite their strong results on platforms with rich profiles and click logs, they depend heavily on high-quality, unbiased user features and thus do not transfer well to long-tail search scenarios.

\begin{figure*}[t]
  \centering
  \includegraphics[width=1.02\linewidth]{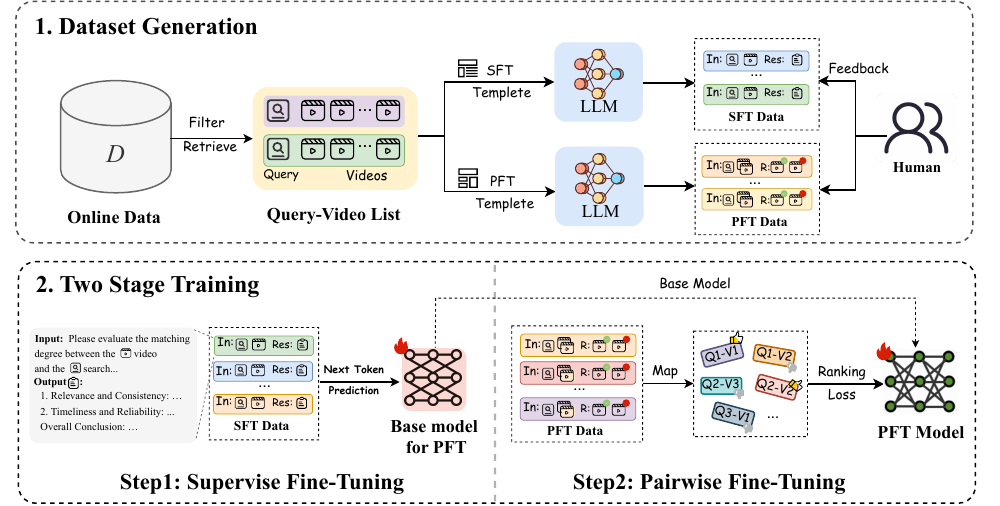}
  \caption{Data construction and two-stage training of the multimodal experience-scoring model, including supervised fine-tuning with multimodal annotations and pairwise fine-tuning with preference data.}
  \label{fig:method-overview}
\end{figure*}

\subsection{Multimodal Alignment and Synthetic Data}

In rich-media retrieval, textual, visual, and audio cues often misalign, causing click-bait or content mismatch; meanwhile, long-tail items suffer from sparse labels. Recent work tackles these issues via multimodal alignment and synthetic supervision.
RagVL\cite{chen2024mllm} fine-tunes a vision-language model as a reranker and injects visual noise during training to boost robustness for image retrieval.
Google Multimodal Reranking\cite{wen2024multimodal} for Knowledge-Intensive VQA fuses vision, text, and knowledge vectors through cross-attention, improving both ranking metrics and VQA accuracy. 
In industry, LLM-Alignment Live-Streaming\cite{liu2025llm} compresses frames, ASR transcripts, and live comments into a unified embedding to filter modality-inconsistent streams under strict latency budgets.
When human labels are scarce, generative models can create auxiliary signals offline. Promptagator\cite{dai2022promptagator} uses a handful of seed examples to mass-generate queries, yielding dense retrievers and rerankers that outperform fully supervised baselines. 
EnrichIndex\cite{chen2025enrichindex} enriches each document with LLM-generated summaries and Q\&A pairs, building a semantically enhanced index that boosts recall and nDCG without extra online inference. 
Doc2query-style pseudo-query expansion\cite{nogueira2019doc2query, li2025semi} is likewise widely adopted.
Collectively, multimodal consistency checks and synthetic labels provide the data and modeling capacity needed for sparse, multimodal long-tail retrieval; this requirement is the core motivation for the two-stage training strategy proposed in our work.

%% file: main_text/method.tex
\begin{figure*}[t]
  \centering
  \includegraphics[width=1.02\linewidth]{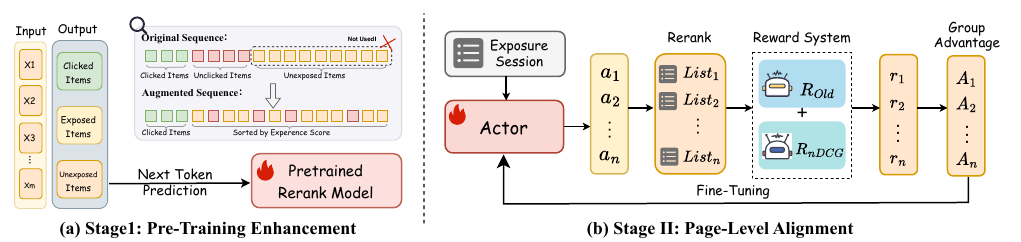}
  \caption{Overview of the two-stage reranking integration. Stage I enhances pre-training with experience-score-ordered supervision; Stage II aligns page-level ranking through GRPO-based reward optimization guided by experience-derived nDCG.}
  \label{fig:method-overview2}
\end{figure*}

\subsection{Framework Overview}

To mitigate long-tail ranking bias in short-video search, we propose an experience score–driven multimodal reranking framework that decouples \emph{score learning} from \emph{ranking deployment}. Specifically, we first train a experience-scoring model offline to predict a experience score for each query-video pair; we then inject this score as an additional, behavior-agnostic signal into the production reranking pipeline to adjust point-wise ordering and to guide page-level policy optimization.

The overall framework consists of three main stages, where the first two are illustrated in Figure \ref{fig:method-overview} and the third stage in Figure \ref{fig:method-overview2}. First, in the mutimodal quality alignment stage, we construct a high-quality multimodal annotation dataset. A large language model is prompted to generate and align dimension-wise quality analyses, enabling the model to learn to evaluate content using textual, speech, and visual signals.
Second, in the pairwise ranking alignment stage, we introduce intra-query video preference pairs and train the model with a pairwise ranking loss. This allows the model to learn partial orderings among candidates, thereby improving score comparability and mitigating the noise introduced by popularity bias and cross-modal mismatch in behavior-driven models.
Finally, in the Integration into the Ranking Pipeline stage, the learned  scores are integrated into the production ranking system. Point-wise scores are used to promote high-quality but under-exposed candidates, while sequence-level scores serve as reinforcement learning rewards to optimize page-level ranking strategies and enhance overall user experience.

\subsection{Multimodal Quality Alignment}

\tcbset{
  colback=black!1,
  % colframe=black!20,   % border color
  arc=10pt,            % corner radius
  boxrule=1pt,           % border width
  float,                 % 让盒子变成可浮动对象（类似 figure）
  floatplacement=tbp,    % 浮动位置建议
  left=10pt,right=10pt,top=8pt,bottom=8pt, % padding
}

\subsubsection{Dataset.} 

Existing open datasets focus mainly on text-only or vision-only retrieval and lack resources that simultaneously cover text, speech, and visual signals for long-tail short-video search. Consequently, models cannot directly learn fine-grained judgments of cross-modal consistency or multi-dimensional content quality. To bridge this gap, we design an LLM-based multi-dimensional annotation pipeline. After obtaining the annotated data, we use it to fine-tune the backbone LLM, effectively enhancing the model's multimodal quality perception and explanatory capabilities.

Specifically, we begin by mining search logs from the past month and filtering out long-tail queries whose seven-day page view (PV) count is below 70. After removing noisy queries, we randomly sample a subset of long-tail queries and retrieve up to ten candidate videos for each query, forming a collection of ⟨query, video⟩ pairs.
For each video, we collect multimodal inputs, including textual features (title, description, OCR snippets, and the full ASR transcript) and visual features (the cover image and four key frames).
Leveraging the comprehension and generation capabilities of existing multimodal models, we design a prompting template that guides the model to produce dimension-wise reasoning analyses for each sample across multiple axes, such as relevance and consistency, image safety and age appropriateness, as well as timeliness and credibility.
To ensure annotation reliability, we manually spot-check a subset of samples, analyze the causes of errors, and iteratively refine the evaluation dimensions in the prompt. This annotation process yields multidimensional quality assessment data, which serve as supervision signals for subsequent model fine-tuning.

\subsubsection{Supervised Fine Tuning}

We cast multimodal quality assessment as a standard next-token prediction task for an LLM. Given a multimodal prompt $x$, which concatenates a fixed system prefix, the user query, and fused textual–audio–visual descriptions, the model is trained to autoregressively generate a complete quality analysis $y$. Let $y$ consist of $T$ tokens; the conditional likelihood factorises as
\begin{equation}
\label{eq:ntp}
p_{\theta}(y \mid x)=\prod_{t=1}^{T}
      p_{\theta}\!\bigl(y_{t}\mid y_{<t},x\bigr).
\end{equation}
The training objective minimises the negative log-likelihood
\begin{equation}
\label{eq:loss}
\mathcal{L}_{\mathrm{SFT}}
  = -\!
      \underset{\langle x,y\rangle\in\mathcal{D}_{\mathrm{SFT}}}
      {\mathbb{E}}
      \Bigl[
        \sum_{t=1}^{T}
        \log p_{\theta}\!\bigl(y_{t}\mid y_{<t},x\bigr)
      \Bigr],
\end{equation}
where $\mathcal{D}_{\mathrm{SFT}}$ is built by prompting LLM
to produce multiple dimension analyses for each \(\langle\text{query},\text{video}\rangle\) pair.
Because the system-prefix tokens are identical across samples,
their prediction losses are masked out during optimisation,
and only the analysis portion contributes to
\(\mathcal{L}_{\mathrm{SFT}}\). After this stage, the model acquires the ability to assess content quality from multiple perspectives conditioned on multimodal evidence.

\subsection{Pairwise Ranking Alignment}

\subsubsection{Dataset.}

In the pairwise ranking alignment stage, we construct multiple candidate video pairs 
\(\langle d_i, d_j\rangle\) for each long-tail query \(q\) based on its exposed videos.  
A comparison-style prompt template is designed to present the multimodal information of both videos 
to the large language model and to request a preference decision according to the quality dimensions 
defined in the first stage.  
This process produces comparable pairwise preference annotations, 
enabling the model to learn the relative quality ranking of videos under the same query.  
To ensure the accuracy and consistency of the annotations, 
we perform multi-level quality control after labeling.  
Specifically, we combine rule-based filtering with manual verification 
to identify and correct partial-order conflicts among videos under the same query, 
ensuring logical consistency and overall reliability of the final preference labels.

\subsubsection{Pairwise Preference Fine Tuning.} 

Building upon the SFT model, we replace the generative head with a
\textit{sequence classification head} to output a comparable scalar quality score.
Given a query–video pair $(q,v)$, the multimodal backbone first encodes the fused representation $h_{q,v} = \mathrm{Mo}(q,v)$, and the quality score is obtained through a linear projection:
\begin{equation}
s_{q,v} = f_\theta(q,v) = w^\top h_{q,v} + b,
\end{equation}
where $w$ and $b$ are learnable parameters. For a pair of candidate videos $(A,B)$ under the same query, with annotated preference
$A \succ B$, we denote $s^+ = s_{q,A}$, $s^- = s_{q,B}$.
The model is trained by minimizing the following objective:
\begin{equation}
\mathcal{L} =
-\mathbb{E}\!\left[\log \sigma(s^+ - s^-)\right]
+ \lambda\,\mathbb{E}\!\left[(s^+ + s^-)^2\right],
\end{equation}
where the first term encourages the model to assign higher scores to preferred videos, and the second term regularizes the overall score distribution to remain centered and stable.
Here, $\sigma(\cdot)$ denotes the Sigmoid function, and $\lambda$ is a balancing coefficient. This unified formulation jointly optimizes relative ranking consistency and score stability,
allowing the model to produce comparable point-wise quality scores $f_\theta(q,v)$ for any query-video pair $(q,v)$,
which can be directly applied to downstream reranking and evaluation of long-tail queries. 
Note that in the following, ExpModel refers to the experience-scoring model trained with supervised fine tuning and pairwise preference fine tuning.

% \subsection{APPLICATION}
\subsection{Integration into the Ranking Pipeline}

We integrate the learned experience scores into a two-stage generative reranking pipeline for short-video search to alleviate long-tail bias. 
In \textit{Stage~I -- Pre-training Enhancement}, we rewrite the training target sequence used in the autoregressive next-item prediction objective, aligning it with the order induced by the experience scores. 
This guides the model to prioritize high-quality content during training.
In \textit{Stage~II -- Page-level Alignment}, we construct a page-level consistency signal based on the experience scores and treat it as a label-derived ideal ordering. 
The sequence policy is then optimized through GRPO\cite{shao2024deepseekmath}. 
The reward function combines the existing behavioral objective with a normalized DCG metric computed against the experience-induced ideal list, thereby preserving business performance while improving page-level consistency and user experience.

\subsubsection{Stage I: Pre-Training Enhancement}
For a query $q$, let $\mathcal{S}$ denote the candidate set for a session and $s_{\exp}(q,i)\in[0,1]$ the point-wise experience score of item $i\in\mathcal{C}$. Let $C$ be the set of clicked items, $E$ the set of exposed but not clicked items, and $U=\mathcal{S}\setminus (C\cup E)$ the set of \emph{unexposed} items. We write $\mathrm{sort}(S;\,s_{\exp}\!\downarrow)$ for sorting a set $S$ in descending order by $s_{\exp}$. We construct the training target sequence by concatenating two segments:
\begin{equation}
\label{eq:target}
\mathbf{y}_{\mathrm{aug}}
\;=\;
\big[
\mathrm{sort}(C;\,s_{\exp}\!\downarrow)
\;\triangleright\;
\mathrm{sort}(E\cup U;\,s_{\exp}\!\downarrow)
\big].
\end{equation}
That is, all clicked items are placed in front and internally ranked by $s_{\exp}$, while the exposed-but-not-clicked and unexposed items are merged and jointly sorted by the same score. This preserves the business constraint ``clicked-first'' while allowing high-quality, previously unexposed items to appear early in the supervision sequence.
The loss remains the standard autoregressive next-token (next-item) negative log-likelihood:
\begin{equation}
\label{eq:loss}
\mathcal{L}_{\text{stage1}}
\;=\;
\sum_{t=1}^{|\mathbf{y}_{\mathrm{aug}}|}
-\log \pi_{\theta}\!\big(y_t \mid q,\; y_{<t},\; \mathcal{C}\big).
\end{equation}
No auxiliary terms or structural changes are introduced in Stage~I; only the target sequence is replaced by \eqref{eq:target}.

\subsubsection{Stage II: Page-Level Alignment}

From the learned experience scores, we derive an ideal page-level ranking order that serves as the reference for supervision.  
Let $\mathcal{C}$ denote the candidate set for a given query, and $s_{\exp}(i)$ the experience score assigned to item $i \in \mathcal{C}$.  
We sort the candidates in descending order of their experience scores to obtain the ideal ranking list:
\begin{equation}
\label{eq:ideal}
\mathbf{y}^{s} = \mathrm{sort}\big(\mathcal{C};\, s_{\exp}\!\downarrow\big).
\end{equation}
The position of item $i$ in this ideal list is denoted as $\mathrm{rank}_{\mathbf{y}^{s}}(i)$.  
Based on this order, we define a graded relevance value for each item as:
\begin{equation}
\label{eq:rel}
\mathrm{rel}_{\mathbf{y}^{s}}(i) = K - \mathrm{rank}_{\mathbf{y}^{s}}(i) + 1,
\end{equation}
where $K$ is the maximum list length.  
This assignment ensures that items ranked higher in $\mathbf{y}^{s}$ receive larger scores.  
% Alternatively, one may directly use $s_{\exp}(i)$ as the relevance signal, since both induce the same order.
Given a generated list $\mathbf{y} = (y_1, \dots, y_K)$ produced by the model,  
we compute its discounted cumulative gain (DCG) and normalized DCG (nDCG) with respect to the ideal ranking $\mathbf{y}^{s}$ as:
\begin{equation}
\label{eq:dcg}
\begin{aligned}
\mathrm{DCG}_{@K}\big(\mathbf{y} \mid \mathbf{y}^{s}\big)
&= \sum_{t=1}^{K} \frac{\mathrm{rel}_{\mathbf{y}^{s}}(y_t)}{\log_2(1+t)},\\[3pt]
\mathrm{nDCG}_{@K}\big(\mathbf{y}, \mathbf{y}^{s}\big)
&= \frac{\mathrm{DCG}_{@K}\big(\mathbf{y} \mid \mathbf{y}^{s}\big)}
        {\mathrm{DCG}_{@K}\big(\mathbf{y}^{s} \mid \mathbf{y}^{s}\big)}.
\end{aligned}
\end{equation}
Here, the numerator measures the quality of the generated ranking, while the denominator normalizes the score using the ideal order,  
thus constraining $\mathrm{nDCG}_{@K}$ within the range $[0,1]$.

To align the model’s generation policy with this page-level consistency,  
we design a GRPO-style sequence-level reward that interpolates the existing behavioral objective with the experience-based page utility:
\begin{equation}
\label{eq:reward}
r(q, \mathbf{y}) = 
\alpha\, R_{\mathrm{old}}(q, \mathbf{y})
+ \beta\, \mathrm{nDCG}_{@K}\big(\mathbf{y}, \mathbf{y}^{s}\big),
\end{equation}
where $q$ denotes the search query, $R_{\mathrm{old}}(q, \mathbf{y})$ represents the pre-existing reward computed according to business rules or models,  
and $\alpha, \beta > 0$ are mixing coefficients that control the trade-off between behavioral and page-level objectives.  
The policy $\pi_{\theta}$ is optimized using GRPO updates over trajectory-level returns defined by Eq.~\ref{eq:reward}.  
Intuitively, the first term preserves behavioral performance,  
while the second term drives the policy toward consistency with the experience-derived ideal ordering (Eq.~\ref{eq:ideal}),  
thereby mitigating click-bias amplification and improving long-tail experience.

%% file: main_text/experiments.tex
This section reports offline experiments only. Under a setup that mirrors real short-video search and emphasizes long-tail queries, we fix recall and ranking and evaluate only the reranking stage. We then present the experimental setup and results: first verifying the effectiveness of the two-stage experience-scoring model, then assessing its deployment effect in reranking, followed by minimal ablation analyses as needed.

\subsection{Experimental Setup}

\subsubsection{Dataset}

Before model training, we constructed a large-scale annotated dataset with strict temporal splitting and rule-based deduplication to ensure the reliability and reproducibility of our experiments.  
All supervised and preference annotations are generated offline using Qwen2.5-VL-32B model.
During the Supervised Fine-Tuning (SFT) stage, we collected and generated approximately 187{,}000 \textless Query, Video\textgreater{} samples from search logs within an earlier time window, covering 31{,}392 long-tail queries with an average of 5.96 videos per query.  
Long-tail queries are defined as those with fewer than 70 page views within any consecutive seven-day window.  
The training, validation, and test sets are strictly disjoint at both the query and video levels, with all queries normalized before splitting.  
In the subsequent Pairwise Preference Fine-Tuning (PFT) stage, we further expanded the dataset to 348{,}000 cleaned preference pairs, spanning 42{,}308 long-tail queries with an average of 8.22 pairs per query.

\begin{table}[h]
  \centering
  \caption{Statistics of the Training Datasets Used for SFT and PFT Stages.}
  \label{tab:dataset_stats}
  \begin{tabular}{@{}cccc@{}}
    \toprule
    \textbf{Dataset} & \textbf{Sample Size} & \textbf{Queries} & \textbf{Avg. per Query} \\
    \midrule
    SFT Data & 187,150 & 31,392 & 5.96 \\
    PFT Data & 347,935 & 42,308 & 8.22 \\
    \bottomrule
  \end{tabular}
\end{table}

\subsubsection{Metrics}

We evaluate the proposed experience score model and its downstream reranking framework entirely through offline metrics that capture ranking accuracy, listwise quality, and perceptual consistency. Specifically, we first assess the model’s intrinsic ability to distinguish preferred items using Pairwise Accuracy (Acc) or AUC on a balanced preference test set, which measures whether the model consistently assigns higher scores to preferred videos; under this balanced setup, PairAcc is equivalent to AUC and is reported as accuracy (\%). We further employ NDCG@K as the primary metric, to evaluate listwise ranking quality by verifying whether high-quality items are ranked toward the top. In addition, a human preference evaluation (GSB) provides a perceptual measure of quality, where annotators compare the model-generated ranking against a baseline as Good, Same, or Bad, and we summarize the advantage rate to quantify net human preference. Beyond evaluating the model itself, we also examine its contribution when incorporated into the two-stage reranking pipeline, applying the same offline metrics on the reranked results. This setup allows us to determine whether the experience model improves overall ordering consistency and page-level quality. Together, these evaluations provide a comprehensive view of the model’s capacity to produce human-aligned experience scores and its effectiveness as a scoring component in the reranking stage.

\begin{table}[t]
  \centering
  \caption{Comparison of baseline methods on the human-labeled query set: NDCG@\{1,5,10\}.}
  \label{tab:ndcg-results}
  \begin{tabular}{cccc}
    \toprule
    \textbf{Method} & \textbf{NDCG@1} & \textbf{NDCG@5} & \textbf{NDCG@10} \\
    \midrule
    GPT-4o (T-P)  & 0.793 & 0.824 & 0.905 \\
    GPT-4o (VL-P) & 0.811 & 0.837 & 0.921 \\
    GPT-4o (T-L)  & 0.687 & 0.752 & 0.885 \\
    RankGPT       & 0.759 & 0.801 & 0.904 \\
    BGE-m3        & 0.612 & 0.721 & 0.864 \\
    ExpModel(Ours)          & \textbf{0.849} & \textbf{0.854} & \textbf{0.930} \\
    \bottomrule
  \end{tabular}
\end{table}

\subsubsection{Baselines}

We evaluate our method against several categories of reference baselines under a controlled zero-shot setting. The baselines fall into three groups: (i) API-based large language model references, (ii) listwise rerankers based on LLM prompting, and (iii) dual-encoder retrieval models. All methods operate on the same per-query candidate pool with matched input budgets, including truncated ASR/OCR text, a fixed number of visual frames or cover images, and identical evaluation lists. This ensures that performance differences reflect genuine ranking quality rather than variations in recall. All external LLMs are kept strictly in zero-shot mode. This zero-shot configuration mirrors realistic deployment settings of API-based models, providing a capacity-oriented reference rather than a task-optimized competitor.

\textbf{API Models.} For the API-based LLM references, we employ several GPT-4o\footnote{\url{https://platform.openai.com/docs/models/gpt-4o}} configurations to measure capacity-oriented zero-shot performance. 
Specifically, \textbf{GPT-4o (T-P)} denotes the text-based pointwise setting, where GPT-4o receives each query along with the candidate’s textual metadata and outputs a scalar score for that pair; the ranked list is obtained by sorting candidates in descending order of these scores. \textbf{GPT-4o (T-L)} refers to the text-based listwise variant, which processes the entire candidate set within a single prompt and directly outputs a global ranking order. 
\textbf{GPT-4o (VL-P)} indicates the multimodal pointwise variant that extends the same procedure by additionally providing visual cues such as cover images and key frames, serving as an approximate upper bound for zero-shot multimodal ranking. 

\textbf{Listwise Rerankers.} The listwise reranking baseline, RankGPT \cite{sun2023chatgpt} performs zero-shot listwise permutation generation: the LLM is prompted to output an ordering of a group of candidates; due to context limits, RankGPT applies a sliding-window strategy that reranks the tail window first and then moves back-to-first, merging window-level permutations into a final list.

\textbf{Dual-Encoder Retrieval Models.} The dual-encoder BAAI/bge-m3(dense mode)\cite{chen2024bge} represents a traditional retrieval paradigm. It encodes both the query and each candidate’s textual fields, computes cosine similarity between normalized embeddings, and sorts candidates by similarity in descending order to obtain the final ranked list.

\subsection{Evaluation Result}

\subsubsection{Effectiveness of the experience-scoring model.} 

We first assess the ranking capability of the two-stage experience-scoring model (ExpModel) on the human-labeled query set. As shown in Table \ref{tab:ndcg-results}, ExpModel achieves the best NDCG@{1,5,10} scores, outperforming baselines across all cutoffs. The gain @1 is particularly notable, indicating stronger discrimination among top-ranked items and more effective promotion of high-quality, query-consistent results.
Comparisons show that zero-shot GPT-4o variants generalize well but lack task-aligned, explicit quality modeling for short-video search. RankGPT is single-modal and cannot exploit visual evidence, which limits its performance. BGE-m3 emphasizes textual relevance and is less expressive for cross-modal consistency and content quality. Our two-stage approach aligns multimodal evidence, learns pairwise preferences for query-aligned quality, yielding more robust gains across cutoffs, particularly at the top ranks.

\begin{table}[t]
  \centering
  \caption{Performance of integrate the experience scores into two-stage reranking pipeline on the human-labeled query set: NDCG@\{1,5,10\}.}

  \label{tab:rerank_exp}
  \begin{tabular}{cccc}
    \toprule
    \textbf{Model} & \textbf{NDCG@1} & \textbf{NDCG@5} & \textbf{NDCG@10} \\
    \midrule
    Exposure Seq         & 0.707 & 0.754 & 0.884 \\
    CTR score            & 0.610 & 0.687 & 0.854 \\
    Relevance score      & 0.611 & 0.722 & 0.866 \\
    Quality score        & 0.564 & 0.673 & 0.843 \\
    Base rerank           & 0.763 & 0.794 & 0.891 \\
    Base + S1       & 0.782 & 0.816 & 0.903 \\
    Base + S1\&S2    & 0.785 & 0.822 & 0.910 \\
    ExpModel          & \textbf{0.849} & \textbf{0.854} & \textbf{0.930} \\
    \bottomrule
  \end{tabular}
\end{table}

\subsubsection{Effectiveness of reranking deployment.} 

To assess the effect of injecting the two-stage experience score into reranking on user experience and consumption behavior, we conduct two complementary experiments using Tables \ref{tab:rerank_exp} and \ref{tab:rerank_cons}. First, in Table \ref{tab:rerank_exp}, we keep recall and ranking fixed and vary only the reranking signals and strategies: \textit{Base}, \textit{Base+S1} which adds the pre-training enhancement, and \textit{Base+S1\&S2} which adds the page-level alignment; we report listwise experience metrics such as NDCG@\{1,5,10\} to test whether the pre-training enhancement and the page-level alignment yield perceptible top-of-page gains. Second, in Table \ref{tab:rerank_cons}, under the same reranking settings, we evaluate business-critical consumption proxies, including long-play ratio and click metrics, to determine whether experience optimization produces only controllable substitution effects on consumption behavior.

In Table \ref{tab:rerank_exp}, when the reranking side is progressively enhanced from \textit{Base} to \textit{Base+S1} and \textit{Base+S1\&S2}, the experience metrics exhibit a monotonic improvement. This indicates that the two-stage training yields a comparable and de-biased pointwise experience score that consistently pushes high-quality, query-consistent content toward more prominent positions, while the page-level alignment further optimizes the relevance and diversity at the top ranks, with particularly pronounced gains in long-tail scenarios. In contrast, the other signals in the table each have limitations: \textit{CTR score} reflects popularity but remains influenced by historical behavior and position/exposure propensity; \textit{Quality score} emphasizes content quality but lacks an explicit constraint for query alignment, so it is difficult to support ranking on its own; \textit{Relevance score} provides strong semantic matching but does not explicitly address exposure/position bias or multimodal inconsistency, and it also lacks page-level governance; \textit{Exposure Seq} denotes the current online ordering visible to users, determined jointly by base rerank and downstream pipeline strategies, and it tends to preserve the existing exposure structure rather than actively upgrading items to the very top. Using the experience model directly for ranking (\textit{ExpModel}) achieves the best overall results, which reflects the upper bound of the two-stage model capacity and validates the effectiveness of injecting its scores into the reranker.

In the offline evaluation of Table \ref{tab:rerank_cons}, with recall and ranking held fixed, the long-play and click metrics for \textit{Base}, \textit{Base+S1}, and \textit{Base+S1\&S2} remain in a similar range. After injecting the experience scores into pre-training lables and adding page-level alignment, we observe only a slight and controllable decline relative to \textit{Base}, with no significant degradation in consumption. This pattern aligns with the gains on the experience side: the page prioritizes higher-quality, query-consistent content, which moderately reduces reliance on historical clicks and exposure tendencies and thus introduces small, acceptable fluctuations in consumption.

\begin{table}[t]
  \centering
  \caption{Evaluation of models on long-play and click-through metrics.}
  \label{tab:rerank_cons}
  \begin{tabular}{ccccc}
    \toprule
    \multirow{2}{*}{\textbf{Model}} & \multicolumn{2}{c}{\textbf{Long-play}} & \multicolumn{2}{c}{\textbf{Click}} \\
    \cmidrule(lr){2-3}\cmidrule(lr){4-5}
    & \textbf{AUC} & \textbf{NDCG@10} & \textbf{AUC} & \textbf{NDCG@10} \\
    \midrule
    Exposure Seq       & 0.595 & 0.879 & 0.612 & 0.883 \\
    CTR Score               & 0.762 & 0.918 & 0.863 & 0.920 \\
    Relevance score         & 0.587 & 0.797 & 0.609 & 0.805 \\
    Quality score           & 0.572 & 0.829 & 0.663 & 0.830 \\
    Base Rerank              & \textbf{0.696} & \textbf{0.877} & \textbf{0.680} & \textbf{0.898} \\
    Base + S1             & 0.694 & 0.872 & 0.679 & 0.883 \\
    Base + S1\&S2       & 0.691 & 0.866 & 0.675 & 0.879 \\
    ExpModel               & 0.654 & 0.863 & 0.673 & 0.868 \\
    \bottomrule
  \end{tabular}
\end{table}

\begin{figure*}[t]
  \centering
  \includegraphics[width=0.95\linewidth]{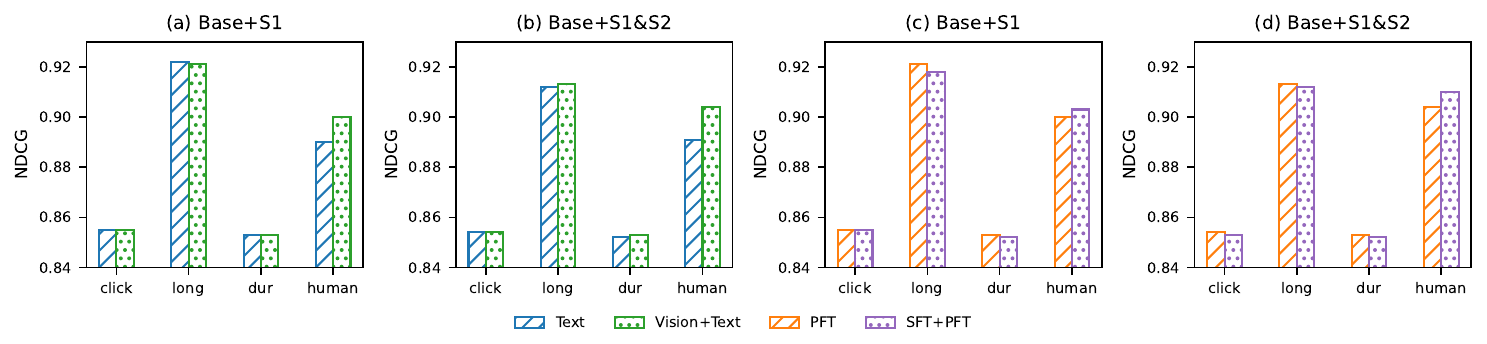}
  \caption{Ablations within the two-stage optimization in reranking. 
Bars report NDCG under four label types:
\textbf{click}: click-based labels, 
\textbf{long} = long-play labels, 
\textbf{dur} = watch-duration labels, 
\textbf{human} = human-preference labels.}
  \label{fig:ablation}
\end{figure*}

\subsubsection{Human Preference Evaluation}

We assess user satisfaction via GSB (Good/Same/Bad) pairwise judgments against the base rerank. As shown in Table \ref{tab:gsb-eval}, using the two-stage experience model directly for ranking (ExpModel) yields a strong net advantage (Adv +11.7\%), representing an upper bound of model capacity. When its scores are injected into the production reranker with page-level alignment (Base + S1\&S2), a clear positive advantage remains (Adv +5.5\%) with a stable proportion of “Same,” indicating that the model’s scoring ability translates into tangible page-level gains, while the magnitude is moderated by deployment constraints. This aligns with our offline metrics: experience gains are stable and the perturbation to behavior is controlled.

\begin{table}[htbp]
  \centering
    \caption{Human GSB evaluation vs. base rerank. 
    Annotators compare each method’s top-$K$ list with the base and label Good/Same/Bad. 
    \textbf{Adv} = $(G-B)/(G+S+B)\times100\%$.}
  \label{tab:gsb-eval}
  \setlength{\tabcolsep}{8pt}
  \begin{tabular}{ccccc}
    \toprule
    \textbf{Method} & \textbf{Good} & \textbf{Same} & \textbf{Bad} & \textbf{Adv} \\
    \midrule
    ExpModel  & 48 & 114 & 26 & \textbf{+11.70\%} \\
    Base + S1\&S2 & 39 & 133 & 28 & +5.50\% \\
    \bottomrule
  \end{tabular}
\end{table}

\subsection{Ablation study}

\begin{table}[t]
  \centering
  \caption{Accuracy of different training schemes and modalities.}
  \label{tab:two-stage-acc}
  \begin{tabular}{c c c c}
    \toprule
    \textbf{Scheme} & \textbf{Modality} & \textbf{Model} & \textbf{Accuracy (\%)} \\
    \midrule
    PFT                    & text         & Qwen3-4B          & 79.9 \\
    SFT + PFT        & text         & Qwen3-4B          & 80.7 \\
    PFT                    & text+image   & Qwen2.5-VL-3B     & 82.2 \\
    SFT + PFT        & text+image   & Qwen2.5-VL-3B     & 84.6 \\
    \bottomrule
  \end{tabular}
\end{table}

\subsubsection{Impact of Two-stage training}

We first assess the necessity of two--stage training by comparing \emph{pairwise fine–tuning only} with the complete \emph{SFT $\rightarrow$ PFT} pipeline under both text–only and multimodal settings.  
With identical model capacity and data, inserting the supervised fine–tuning stage raises ACC from \(79.9\%\) to \(80.7\%\) on the text model and from \(82.2\%\) to \(84.6\%\) on the multimodal model, demonstrating that evidence-grounded supervision provides a stronger foundation upon which pairwise preference learning can build.

\subsubsection{Impact of Modalities and Training Strategy on Reranking.}
As shown in Figure \ref{fig:ablation} (a) and (b), at either stage the text-only input underperforms the multimodal setting (text + vision), with a larger gap on experience-oriented metrics such as \emph{human}. On behavior-driven proxies (\emph{click}, \emph{long}, \emph{dur}), the difference is smaller. This indicates that visual cues complement textual signals, enhance query–content consistency, and help surface high-quality items at the top ranks. The pattern is consistent in both pre-training enhancement and page-level alignment.
As shown in Figure \ref{fig:ablation} (c) and (d), adding the SFT stage before PFT yields consistent improvements on the \emph{human} metric in both stages, while maintaining performance on behavior-driven proxies at a comparable level. Establishing a comparable, de-biased base score with SFT and then refining the relative order with PFT better captures the notion of “query-aligned quality.” This conclusion holds across both stages.

\section{Online Applications}

\subsection{Deployment Details}

To validate the feasibility and effectiveness of our proposed experience score-driven two-stage reranking framework in real-world search scenarios, we deployed the method in Kuaishou’s production search system and conducted a large-scale A/B test. The experiment targeted the long-tail query segment, defined as queries with fewer than 70 page views over a 7-day window. We sampled 5\% of total search traffic for the test bucket, within which long-tail queries accounted for approximately 15.4\%. The experiment lasted two weeks and handled over 50 million queries per day, ensuring statistical significance and broad representativeness.

The reranking model was implemented in TensorFlow, with a compact 4M parameter architecture, enabling efficient online inference and deployment. Online inference was performed on 2 nodes, each equipped with 2 NVIDIA A10 GPUs. For training, the model was optimized using our constructed multimodal long-tail annotation dataset, on a distributed setup with 6 nodes, each using 2 Tesla T4 GPUs (16 GB). The first stage employed supervised fine-tuning, with a per-node batch size of 1024, using the AdamW optimizer and a learning rate of $5 \times 10^{-6}$. The second stage incorporated a page-level reinforcement learning strategy based on GRPO, using 64-sample rollouts and the same learning rate. In practice, $\alpha$ and $\beta$ were fixed business constraints to ensure online stability while balancing the two reward signals.
The training pipeline followed a batched update schedule, with model weights refreshed every 12 hours to adapt to evolving data distributions.

\subsection{Performance of A/B Test}

As shown in Table~\ref{tab:exp4_metrics}, the online A/B test results confirm the effectiveness of our experience score–driven two-stage reranking framework in improving both user experience and ranking quality. In the long-tail query segment, all three core metrics show significant gains: the IQRR rate dropped by 1.28\%, indicating that users were more satisfied with the first-page results and less likely to reformulate their queries; the CTR increased by 1.24\%, suggesting better surfacing of relevant candidates; and the LVR rose by 1.67\%, reflecting higher content quality and user retention. These improvements demonstrate that our model can reliably identify and promote high-quality videos even in low-feedback scenarios.
Importantly, these benefits also generalized to overall traffic. Across all queries, the IQRR rate decreased by 0.11\%, while CTR and long-view ratio increased by 0.13\% and 0.19\%, respectively. Although the absolute gains are smaller, they remain statistically meaningful given that long-tail queries account for only 15.4\% of total volume. This suggests that improving long-tail ranking not only enhances sparse-query performance but also contributes to global ranking stability without disrupting consumption behavior. Notably, the increase in online consumption metrics is not at odds with the slight declines observed in offline evaluation: offline labels are derived from historical logs with a fixed page composition and thus measure a static substitution effect. After deployment, the experience score places higher-quality, query-consistent results more prominently, which dynamically reshapes users’ browsing paths and viewing intent, thereby translating into a natural overall rise in consumption.

\begin{table}[t]
\centering
\caption{Online A/B test results on Kuaishou Search. “All queries” refers to the entire test bucket; “Long-tail queries” are defined as those with 7-day PV < 70. \textbf{IQRR} denotes the \emph{Intent Query Reformulation Rate}, \textbf{CTR} is the \emph{Click-Through Rate}, and \textbf{LVR} is the \emph{Long-View Ratio}.}
\label{tab:exp4_metrics}
\label{tab:exp4_metrics}
\begin{tabular}{cccc}
\toprule
\textbf{Metric Name} & \textbf{IQRR} & \textbf{CTR} & \textbf{LVR} \\
\midrule
all queries        & -0.11\% & +0.13\% & +0.19\% \\  
long-tail queries  & -1.28\% & +1.24\% & +1.67\%  \\
\bottomrule
\end{tabular}
\end{table}

%% file: main_text/conclusion.tex
This paper presents an unbiased multimodal reranking framework for long-tail short-video search. By combining multimodal evidence–aligned supervised fine-tuning with pairwise preference optimization, the model learns a comparable, de-biased experience score capturing query alignment and content quality. Injecting this score into the production reranker, together with pointwise enhancement and page-level alignment, improves top-of-page experience metrics without disrupting consumption behavior. Offline results show consistent gains and confirm the model’s ability to identify high-quality content. Large-scale online A/B tests further validate the deployability and scalability of the framework, achieving statistically significant improvements in both user experience and consumption behavior.
For future work, we plan to further refine the two-stage training mechanism by exploring direct listwise modeling to better capture holistic ranking dependencies. We also aim to incorporate richer audio semantics and session-level feedback, extending the framework’s generalization and stability to more diverse, multimodal, and long-tail retrieval scenarios.

%% file: main_text/appen.tex
\section{Prompt Template}

\subsection{Prompt Template for SFT Data Generation.}

\ \ \ \ \textbf{Task}: Judge the match between the search query and the video, and the overall viewing experience, using the provided evidence (query, title, cover, key frames, ASR, OCR).

\medskip
\textbf{Instructions}:\\
Analyze each dimension below. If evidence is missing or inconclusive, write \emph{``Insufficient evidence''} and make a conservative judgment. Keep reasoning concise and tied to the cited evidence (e.g., refer to title/ASR/OCR/frame).

\medskip
\textbf{Evaluation dimensions \& rules}:\\
- Relevance \& Consistency: Check whether title, cover, key frames, and ASR consistently address the query intent...\\
- Image Safety \& Age-Appropriateness: Inspect cover and key frames for lowbrow/violent/gory/sexual/illegal elements or suggestive thumbnails that mislead...\\
- Timeliness \& Trustworthiness: If the query is time-sensitive, judge whether content appears outdated or conflicts with common facts...\\
- xxx: ...

\medskip
\textbf{Output format}:\\
- Relevance \& Consistency: xxx\\
- Image Safety \& Age-Appropriateness: xxx\\
- Timeliness \& Trustworthiness: xxx\\
- xxx...\\
- Overall verdict: Summarize the match between the query and the video and the expected user experience.

\medskip
\textbf{Input}:\\
-- Search query: \{search\_term\}\\
-- Video info: \{video\_info\}
% \end{tcolorbox}

\subsection{Prompt Template for PFT Data Generation.}

\ \ \ \ \textbf{Task}: 
Given a search query and two candidate videos (A and B), compare their overall match to the query and expected user experience, using all available multimodal evidence (title, cover, key frames, ASR, OCR). 
Decide which video provides a better experience for the query.

\medskip
\textbf{Instructions}:\\
Analyze both videos along the following dimensions, referencing specific evidence when possible (e.g., title, ASR/OCR content, visual cues). 
If evidence is missing or unclear, write \emph{``Insufficient evidence''} and make a conservative judgment.
Keep reasoning concise and comparative.
At the end, provide a clear verdict choosing the preferred video.

\medskip
\textbf{Evaluation dimensions \& rules}:\\
- Relevance \& Consistency: Check whether title, cover, key frames, and ASR consistently address the query intent...\\
- Image Safety \& Age-Appropriateness: Inspect cover and key frames for lowbrow/violent/gory/sexual/illegal elements or suggestive thumbnails that mislead...\\
- Timeliness \& Trustworthiness: If the query is time-sensitive, judge whether content appears outdated or conflicts with common facts...\\
- xxx: ...

\medskip
\textbf{Output format}:\\
- Overall Experience: [A better / B better / Tie] + brief reasoning\\
- Final Verdict: Preferred video = A / B.

\medskip
\textbf{Input}:\\
-- Search query: \{search\_term\}\\
-- Video A info: \{video\_A\_info\}\\
-- Video B info: \{video\_B\_info\}